\documentclass[]{spie}  

\usepackage{amsmath,amsfonts,amssymb}
\usepackage{graphicx}
\usepackage[colorlinks=true, allcolors=blue]{hyperref}
\usepackage{subfig}

\title{The development of an optical design tool for atmospheric dispersion correction}

\author[a,b]{Wehbe, B.}
\author[c,d]{Cabral, A.}
\author[e]{\'{A}vila, G.}
\affil[a]{Instituto de Astrof\'{i}sica e Ci\^{e}ncias do Espa\c{c}o, Universidade do Porto, CAUP, Rua das Estrelas, 4150-762 Porto, Portugal}
\affil[b]{Departamento de F\'{i}sica e Astronomia, Faculdade de Ci\^{e}ncias, Universidade do Porto, Rua Campo Alegre, 4169-007 Porto, Portugal}
\affil[c]{Instituto de Astrof\'{i}sica e Ci\^{e}ncias do Espa\c{c}o, Universidade de Lisboa, Campus do Lumiar, Estrada do Pa\c{c}o do Lumiar 22, Edif. D, PT1649-038 Lisboa, Portugal}
\affil[d]{Laborat\'{o}rios de \'{O}ptica Lasers e Sistemas; Dep. F\'{i}sica, Faculdade de Ci\^{e}ncias, Universidade de Lisboa Edific\'{i}o C9, Campo Grande, PT1749-016 Lisboa, Portugal}
\affil[e]{European Southern Observatory, Karl-Schwarzschild-Stra{\ss}e 2, 85748 Garching bei M\"{u}nchen, Germany}

\authorinfo{E-mail: bachar.wehbe@astro.up.pt}

\begin{document} 
\maketitle

\begin{abstract}
In ground based astronomical observations, atmospheric dispersion shifts the image of the object at different wavelengths due to the wavelength-dependent index of refraction of the atmosphere. Thus, using an Atmospheric Dispersion Corrector (ADC) is mandatory in order to avoid any wavelength dependent losses. Typical ADC configurations, for high resolution astronomical instruments, are two counter-rotating prisms, a set of, at least, four prisms paired together. With the arrival of large telescopes with higher angular magnification, and spectrographs with higher resolution, the requirements on the dispersion correction are becoming more critical due to the impact on the produced science (e.g. radial velocity precision). We developed an ADC optical design tool in order to select the best set of glasses in terms of residuals, transmission, resulting image quality, Fresnel losses, taking into account the required spectral range and typical atmospheric conditions where the ADC will be working. A demonstration of the capabilities of the tool is presented with the analysis of the impact of different melt data, the effect of different glass Sellmeier coefficients between catalog and measured ones, that can create a difference in the residuals above few tens of milli-arcseconds (mas). The tool allows the investigation of critical steps on the ADC design phase and speeds up the glass selection process critical for the harder requirements of the future instruments/telescopes.
\end{abstract}


\section{INTRODUCTION}
\label{sec:intro}  
\begin{figure}
\centering
\includegraphics[scale=0.25]{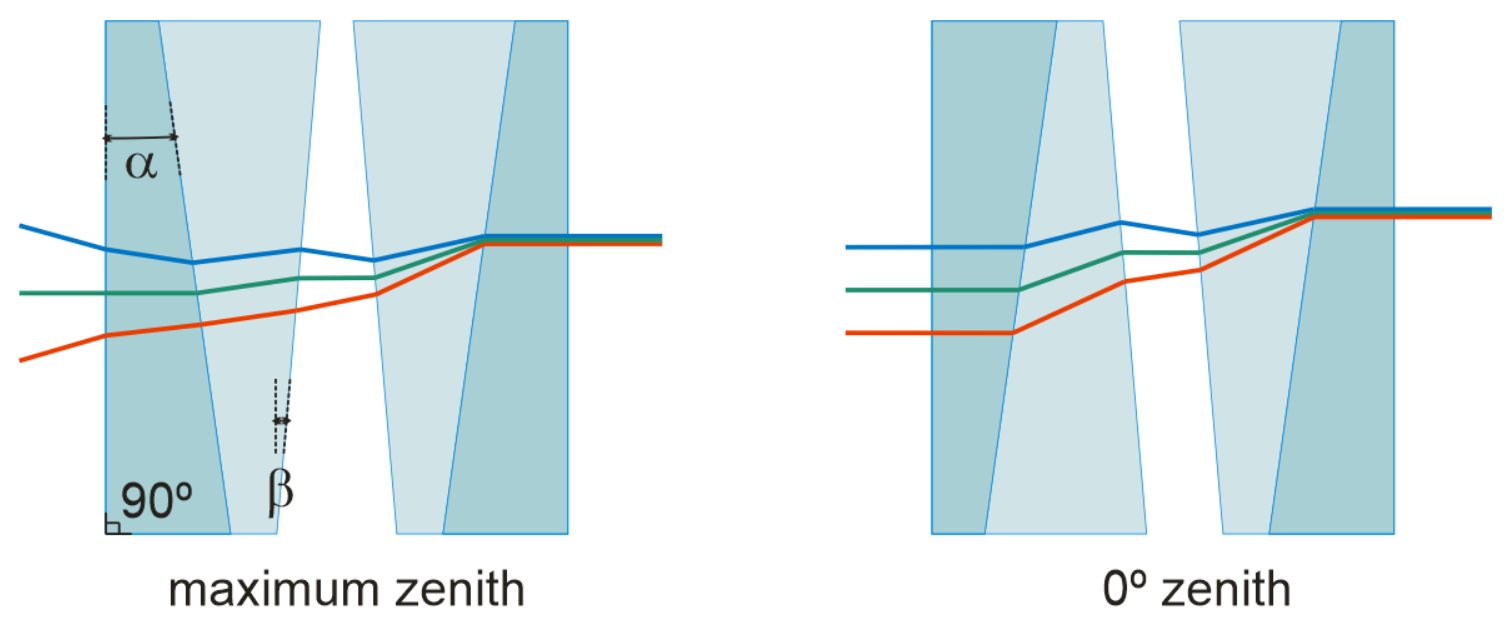}
\caption{ADC with counter rotation prisms, for a maximum zenith (left) and 0$^{\circ}$ zenith (right) configuration \cite{Cabral2012}.}
\label{fig:adc}
\end{figure}

Observations with ground based telescopes are affected by differential atmospheric dispersion due to the wavelength-dependent index of refraction of the atmosphere. The usage of an ADC is fundamental to compensate this effect. The introduction of the ADC in the telescope optical system is a possibility but for large telescopes and high resolution instruments, becomes far from the ideal solution. The inclusion of the ADC in the instrument front end is therefore the current trend. With the arrival of large telescopes with higher angular magnification, and spectrographs with higher resolution, the requirements on the dispersion correction are becoming more critical due to the impact on the produced science (e.g. radial velocity precision \cite{Fischer2016}).
\\
The main requirements for an ADC is to perform variable counter dispersion to compensate that of the atmosphere at a given zenithal angle and to produce zero-deviation at a reference wavelength, within the range of interest for all zenithal angles. Typical ADC configurations to achieve these requirements, when a high level of compensation is needed, are two counter-rotating prisms with dispersion maximum/minimum when the apex angles of the prisms are in the same/opposite directions, respectively. Two prisms are sufficient to satisfy this requirement, but they can't satisfy the zero-deviation condition unless each prism by itself is a zero-deviation unit, a pair of prisms with different dispersion and oppositely directed apex angles. Thus an ADC is a set of, at least, four prisms paired together (see figure \ref{fig:adc}) to satisfy the given requirements.

In order to investigate critical steps on the ADC design phase, speed up the glass selection process and select the best set of glasses in terms of figures of merit, we developed an ADC optical design tool in order to select the best set of glasses in terms of residuals, transmission, resulting image quality, Fresnel losses, taking into account the required spectral range and typical atmospheric conditions where the ADC will be working.

\section{The tool}
\subsection{Description and flow chart}
\begin{figure}[ht]
    \centering
    \includegraphics[scale=0.7]{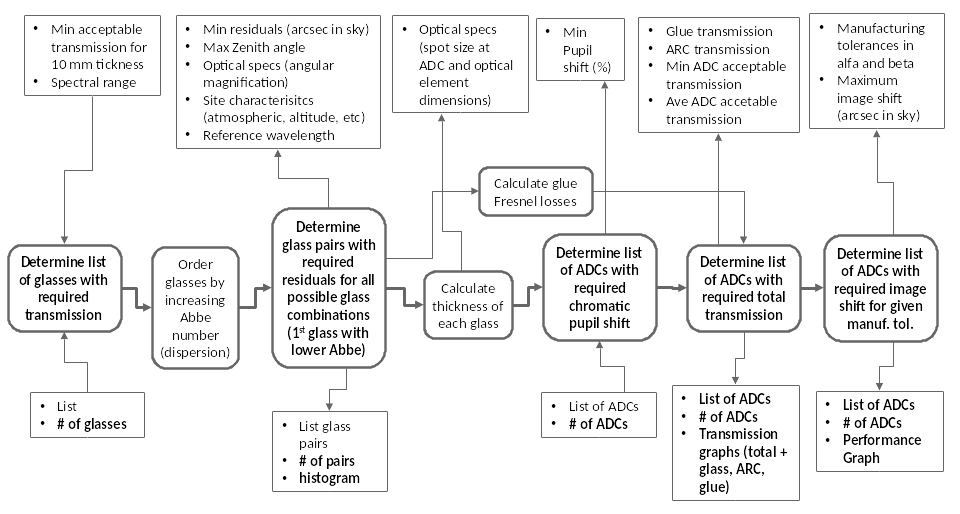}
    \caption{Detailed flow chart of the ADC tool}
    \label{fig:tool}
\end{figure}

As described in the introduction, an ADC is a set of, at least four prisms paired together. In order to test all the possible combinations of glasses, and return the best pair, we made this procedure automated. The different scripts of the tool are developed using Python and then linked to Excel in order to produce a graphical user interface. The tool is described in details in the flow chart (see figure \ref{fig:tool}). We list here its main functions:
\begin{itemize}
    \item Create a glass catalog based on the required minimum transmission over a specified spectral range. The transmission of the glasses is extracted from the catalogs and interpolated over the spectral range of interest.
    \item Setting the telescope parameters (primary diameter, and beam diameter) to get the magnification, and site parameters (temperature, altitude, relative humidity, RH, and zenith angle).
    \item Compute the dispersion residuals between the atmospheric model used (equation \ref{eq:dispersion}) and all the glass combinations taking into consideration the refractive indices of the glasses, the prism angles $\alpha$ and $\beta$ following Snell's law (equation \ref{eq:snell}).
    \begin{align}
    \Delta R(\lambda) &= R(\lambda) - R(\lambda_{ref}) \nonumber  \\
    \Delta R(\lambda)  &\approx 206265 \left[ n(\lambda) - n(\lambda_{ref}) \right] \times \tan z
    \label{eq:dispersion}
    \end{align} 
    where  $\lambda_{ref}$ is the reference wavelength, and $z$ is the zenithal angle of observation.
    
    \begin{equation}
        n_1 \sin \theta_1 = n_2 \sin \theta_2
        \label{eq:snell}
    \end{equation}
    The prisms angles $\alpha$ and $\beta$ are computed following this logic:
    \begin{enumerate}
        \item For each $\alpha$ between 0$^{\circ}$ and 90$^{\circ}$
        \item For each $\beta$ between 0$^{\circ}$ and 90$^{\circ}$
        \item Compute the residuals
        \item Locate $\alpha$ and $\beta$ where we have a minimum of residuals
    \end{enumerate}
    \item Compute the thickness of the glasses of each combination.
    \item Compute the Fresnel losses due to the glue.
    \item Compute the full ADC transmission taking into consideration the thickness of the glasses and the glasses coating.
    \item Compute the chromatic beam shift introduced by the ADC relative to $\lambda_{ref}$.
    \item Create a list of all the ADCs that satisfy the requirements (residuals, transmission, beam shift).
    \item Re-compute the residuals and beam shift of the selected ADC taking into consideration the manufacturing errors on the glasses angles.
\end{itemize}

\subsection{Output example}
\begin{figure}[ht]
\centering
\resizebox{\hsize}{!}{\includegraphics{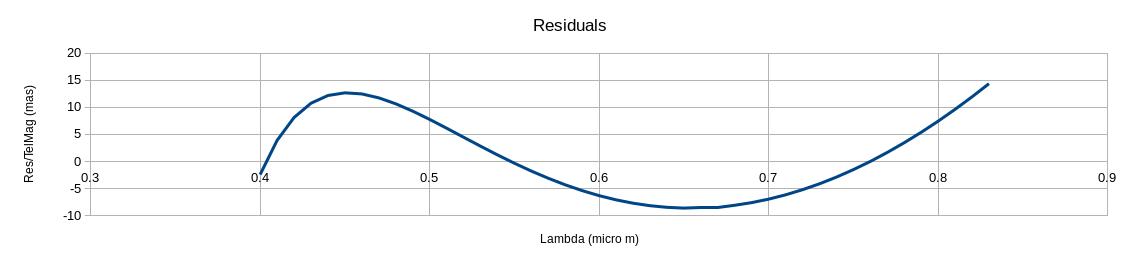}}
\\
\resizebox{\hsize}{!}{\includegraphics{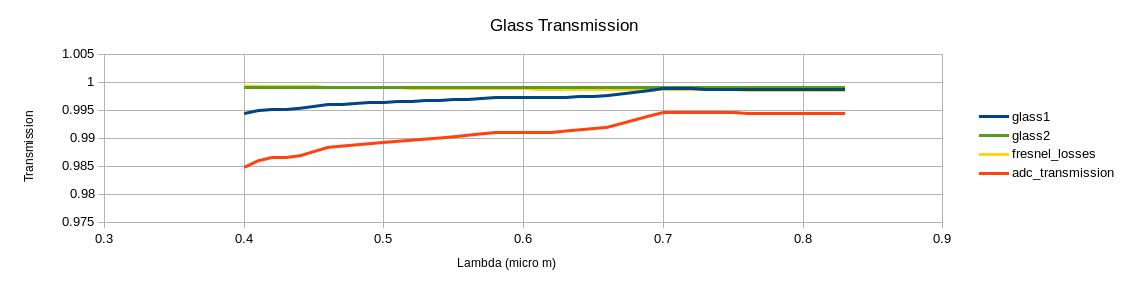}}
\caption{K10 + SFPL51-Y combination results. Top: dispersion residuals; bottom: glasses transmission (glass 1 and 2 are K10 and S-FPL51Y, respectively) and Fresnel losses.}
\label{fig:results}
\end{figure}

The tool creates a list of different ``best" ADC combinations, where the user can easily analyze the graphs and outputs of each combination. The tool now tests all the glasses from Schott \cite{Schott}, Ohara \cite{Ohara} and Nikon. Any extra glass catalog can be easily added. It takes the tool almost 30 minutes to return the results of 45000 different combinations. In figure \ref{fig:results}, we show an example from the several ADC combinations.

The tool was tested using ESPRESSO, NIRPS and HIRES as case studies. One of the ``best" combinations according to the tool is the one chosen already for ESPRESSO (K7 + S-FPL51Y). The tool also revealed another combination N-BAK2 + S-FPL51Y which also seems a good option that had to be considered. It was a similar result for NIRPS, as the tool returned the combination already chosen with no other special combination. \\
As for HIRES, the tool allowed us to investigate several ADC combinations at the same time and in a much faster way than what was done before. It also returned some combinations that weren't taken into consideration before, and allowed us to re-think about the ADC combination of the future HIRES spectrograph on the ELT. Besides the combination used for ESPRESSO, the tool returned several different combinations to be considered and studied more in details (the K10 + S-FPL51Y in particular which shows better results in terms of transmission and residuals). In order to do so, we tried to get access on different melt data for the glasses of interest from the different catalogs we are testing (described in the next section). 

\section{Melt data issue}
As mentioned above, the melt data can cause a serious problem when designing an ADC. When processing the glass, the optical companies use different melting processes that can vary the Sellmeier coefficients from one melt to another. The Sellmeier coefficients provided in the catalogs used during the designing phase, might not be the same when actually buying the glass for manufacturing. This issue first appeared in the integration phase of ESPRESSO. \\
In order to evaluate the errors introduced by different Sellmeier coefficients, we were able to get some melting certificates from Schott and Ohara for some of the glasses of interest (see table \ref{table:glasses}).

\begin{figure}[ht]%
    \centering
    \subfloat[Same angles]{{\includegraphics[width=8cm]{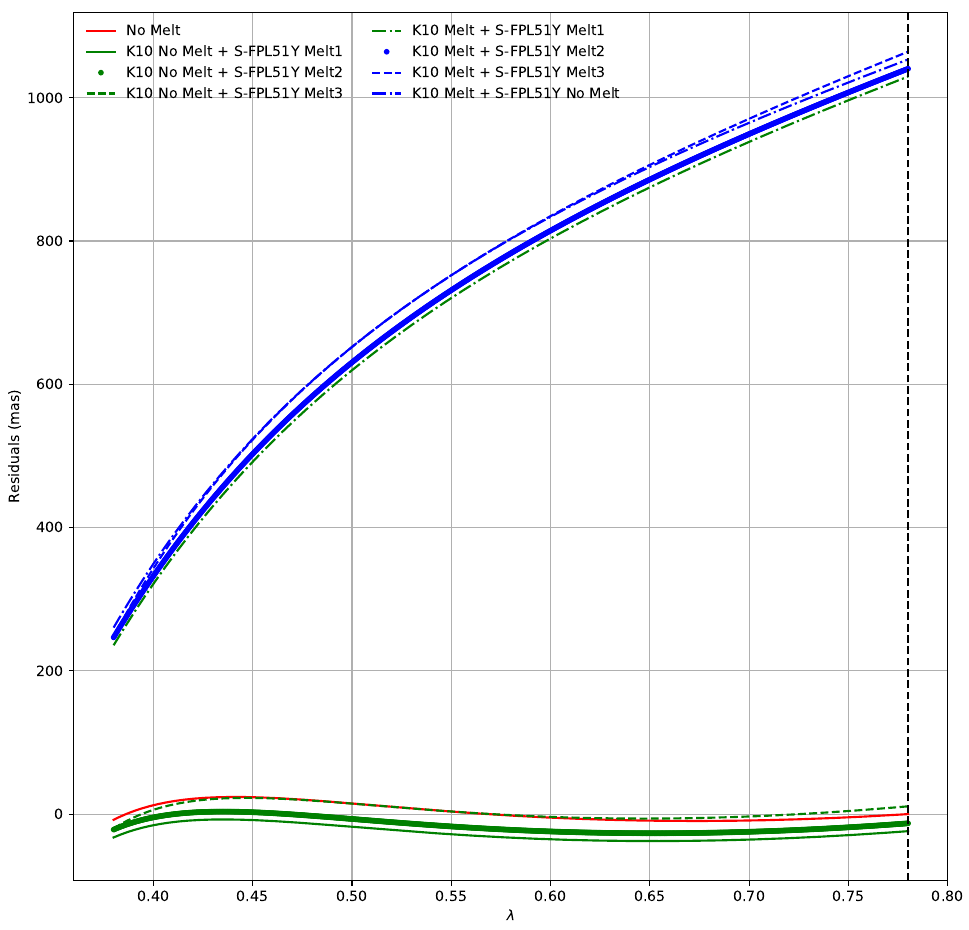} }}%
    \qquad
    \subfloat[Different angles]{{\includegraphics[width=8cm]{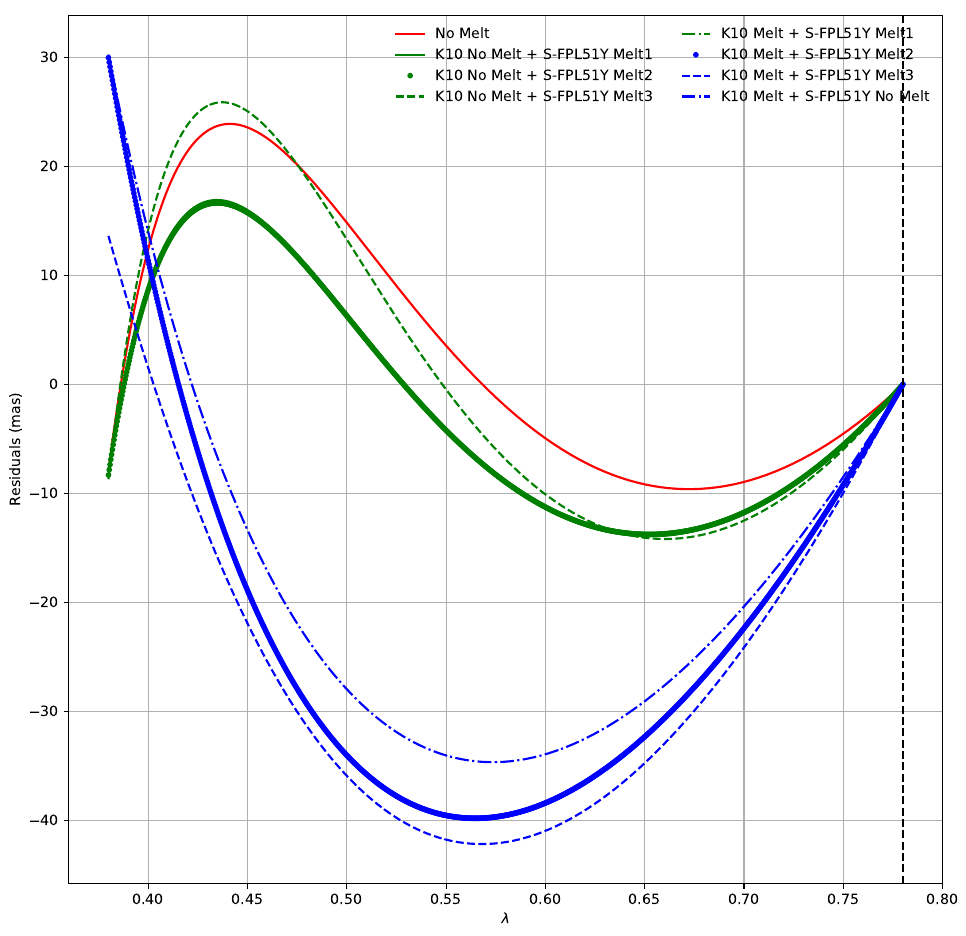} }}%
    \caption{Residuals in mas for all the melt cases of K10 + S-FPL51Y for 2 cases where the prism angles are fixed and when we allowed the change of the angles. The dashed vertical line represent $\lambda_{ref} = 0.78\mu m$}%
    \label{fig:anglesk10}%
\end{figure}

\begin{figure}[ht]%
    \centering
    \subfloat[Same angles]{{\includegraphics[width=8cm]{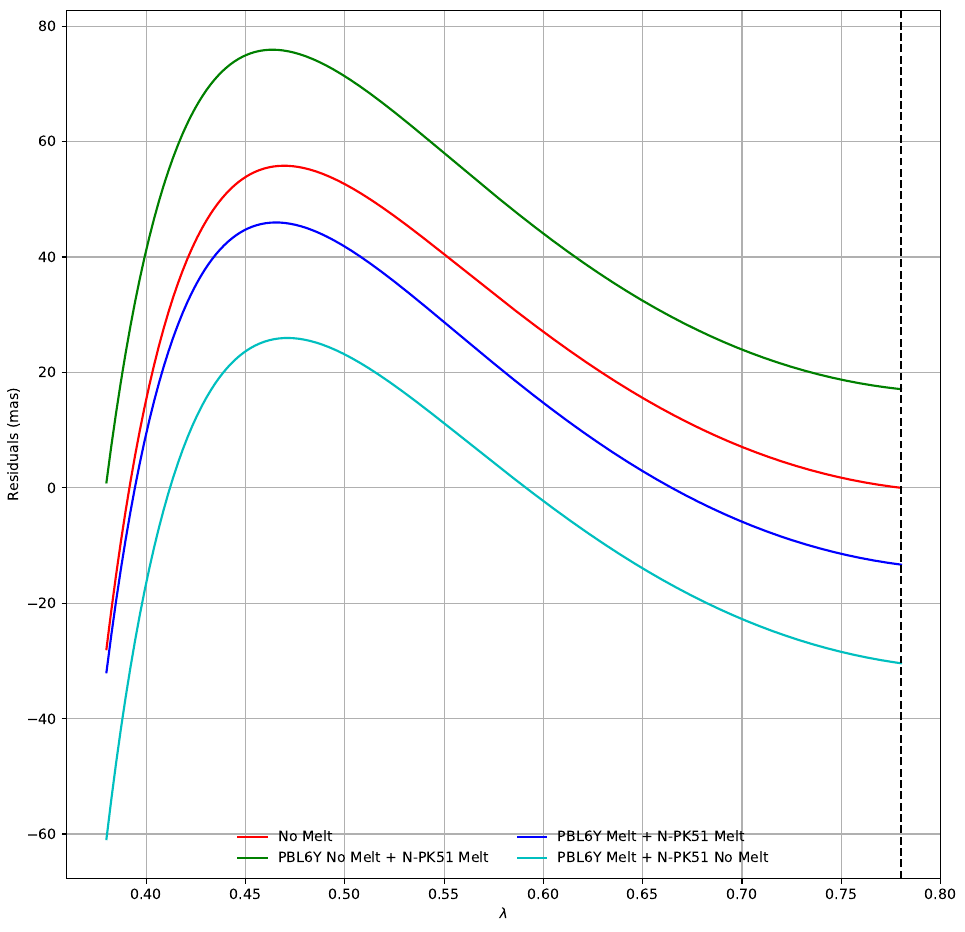} }}%
    \qquad
    \subfloat[Different angles]{{\includegraphics[width=8cm]{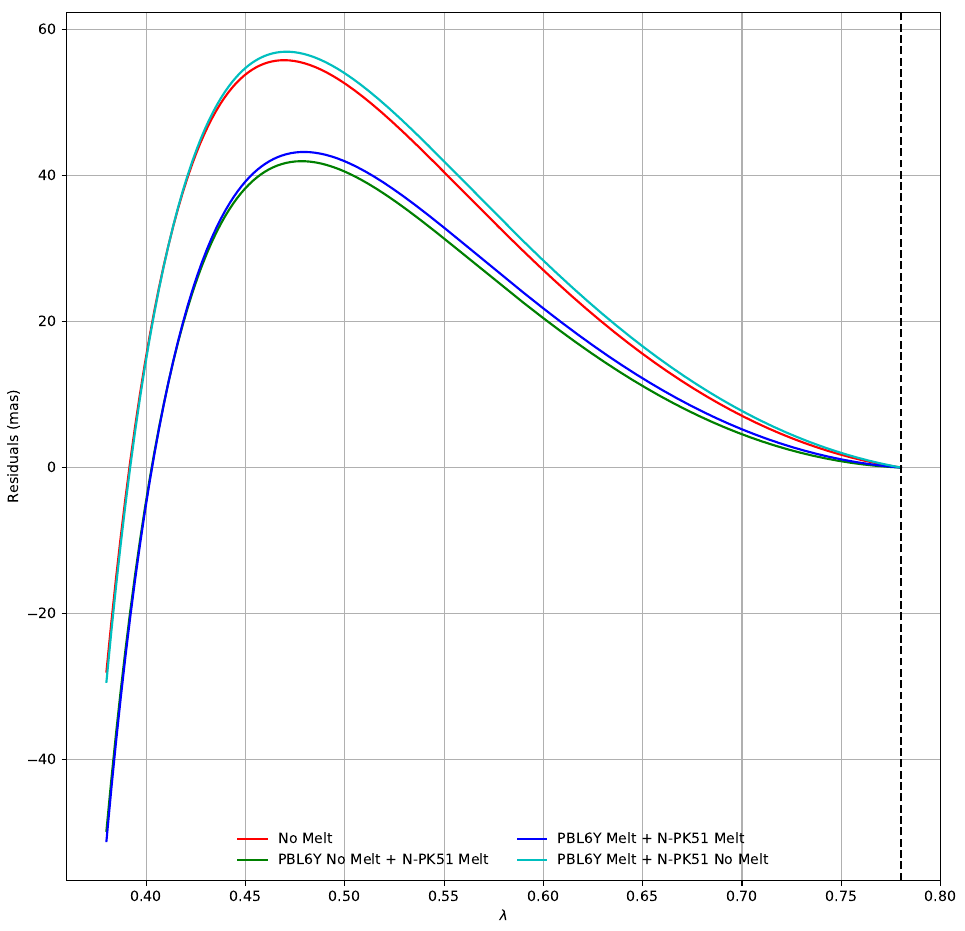} }}%
    \caption{Same as figure \ref{fig:anglesk10} but for PBL6Y + N-PK51.}%
    \label{fig:anglespbl}%
\end{figure}

\begin{table}[ht]
    \centering
    \begin{tabular}{c c}
        Schott & Ohara  \\
        \hline
        K7 & S-FPL51Y \\
        K10 & PBL6Y \\
        N-PK52A & \\
        N-PK51 & \\
        N-FK51A & \\
        \hline
    \end{tabular}
    \caption{Different glasses from Schott and Ohara, that we received their melt certificates.}
    \label{table:glasses}
\end{table}
Using the tool described above, we tested these glasses using different coefficients. Having different melt data, will result in different amounts of residuals, which in some cases might exceed the requirements. We investigated several cases, taking into consideration if we can adjust the prisms angles or not. For example, we will show two cases, K10 + S-FPL51Y in table \ref{table:meltk10} and in figure \ref{fig:anglesk10} and PBL6Y + N-PK51 in table \ref{table:meltpbl} and in figure \ref{fig:anglespbl}. \\
It was clear from the tables and the plots, that the case of K10 + S-FPL51Y was more sensitive to melt data. Even when we allowed the change of prisms angles, the residuals were a lot more different than the catalog (No melt) case. As for the second case, PBL6Y + N-PK51, the results were less sensitive to the melt data variation.

\begin{table}[ht]
    \centering
    \begin{tabular}{c|c c c c}
         Glass & S-FPL51Y & S-FPL51Y MELT1 & S-FPL51Y MELT2 & S-FPL51Y MELT3  \\
         \hline
         K10 & 23.9 & 37.65 & 26.71 & 22.62 \\
         K10 MELT & 1053.27 & 1029.44 & 1040.38 & 1064.13 \\
         \hline
    \end{tabular}
    \caption{Residuals in mas using the same prism angles for different melt combinations of K10 + S-FPL51Y}
    \label{table:meltk10}
\end{table}

\begin{table}[ht]
    \centering
    \begin{tabular}{c|c c}
       Glass  & N-PK51 & N-PK51 MELT \\
       \hline
        PBL6Y & 55.8 & 75.91 \\
        PBL6Y MELT & 60.86 & 45.97 \\
        \hline
    \end{tabular}
    \caption{Same as table \ref{table:meltk10}, but for PBL6Y + N-PK51}
    \label{table:meltpbl}
\end{table}

\section{Conclusion}
The tool allowed us to investigate several ADC combinations at the same time and in a much faster way than what was done before. The tool also returned some combinations that weren't taken into consideration before, and allowed us to re-think about the ADC combination of the future HIRES spectrograph on the ELT. We were also able to investigate the impact of different melt data of the glasses, in terms of designing an appropriate ADC. Since we didn't have a big number of melt data batches, we weren't able to complete a statistical job. What is clear for now, is that people should pay attention and take into consideration the fact that different melt, might return different results that should be taken into consideration.
\\
The tool will be soon available for the community to be used.
 
\acknowledgments 
 
The first author is supported by an FCT fellowship (PD/BD/135225/2017), under the FCT PD Program PhD::SPACE (PD/00040/2012). The authors would also like to acknowledge Ohara, Schott and in particular Mrs Val\'{e}rie Maire for all her help and support in providing the melt data batches.
\newpage
\bibliography{report} 

\begin{thebibliography}{1}

\bibitem{Fischer2016}
{Fischer}, D.~A., {Anglada-Escude}, G., {Arriagada}, P., {Baluev}, R.~V.,
  {Bean}, J.~L., {Bouchy}, F., {Buchhave}, L.~A., {Carroll}, T., {Chakraborty},
  A., {Crepp}, J.~R., {Dawson}, R.~I., {Diddams}, S.~A., {Dumusque}, X.,
  {Eastman}, J.~D., {Endl}, M., {Figueira}, P., {Ford}, E.~B.,
  {Foreman-Mackey}, D., {Fournier}, P., {F{\H u}r{\'e}sz}, G., {Gaudi}, B.~S.,
  {Gregory}, P.~C., {Grundahl}, F., {Hatzes}, A.~P., {H{\'e}brard}, G.,
  {Herrero}, E., {Hogg}, D.~W., {Howard}, A.~W., {Johnson}, J.~A., {Jorden},
  P., {Jurgenson}, C.~A., {Latham}, D.~W., {Laughlin}, G., {Loredo}, T.~J.,
  {Lovis}, C., {Mahadevan}, S., {McCracken}, T.~M., {Pepe}, F., {Perez}, M.,
  {Phillips}, D.~F., {Plavchan}, P.~P., {Prato}, L., {Quirrenbach}, A.,
  {Reiners}, A., {Robertson}, P., {Santos}, N.~C., {Sawyer}, D., {Segransan},
  D., {Sozzetti}, A., {Steinmetz}, T., {Szentgyorgyi}, A., {Udry}, S.,
  {Valenti}, J.~A., {Wang}, S.~X., {Wittenmyer}, R.~A., and {Wright}, J.~T.,
  ``{State of the Field: Extreme Precision Radial Velocities},'' {\em
  pasp}~{\bf 128},  066001 (June 2016).

\bibitem{Cabral2012}
{Cabral}, A., {Moitinho}, A., {Coelho}, J., {Lima}, J., {{\'A}vila}, G.,
  {Delabre}, B.-A., {Gomes}, R., {M{\'e}gevand}, D., {Zerbi}, F., {Di
  Marcantonio}, P., {Lovis}, C., and {Santos}, N.~C., ``{ESPRESSO: design and
  analysis of a Coud{\'e}-train for a stable and efficient simultaneous optical
  feeding from the four VLT unit telescopes},'' in [{\em Ground-based and
  Airborne Telescopes IV}{\nolinebreak\hspace{0.1em}]},  {\em procspie} {\bf
  8444},  84444F (Sept. 2012).

\bibitem{Schott}
SCHOTT,  [{\em {Optical Glass Catalog}}{\nolinebreak\hspace{0.1em}]} (2018).

\bibitem{Ohara}
Ohara,  [{\em {Optical Glass Catalog}}{\nolinebreak\hspace{0.1em}]} (2018).

\end{thebibliography}
\bibliographystyle{spiebib} 

\end{document}